\documentclass[aps,pre,showpacs,showkeys,preprintnumbers,amsmath,amssymb,superscriptaddress,twocolumn,longbibliography,floatfix]{revtex4-2}
\usepackage[english]{babel}
\usepackage{makeidx} 
\usepackage{graphicx} 
\usepackage{dcolumn}
\usepackage{array}
\usepackage{amssymb} 
\usepackage{amsmath}
\usepackage{textcomp}
\usepackage{multirow}
\usepackage{subfigure}
\usepackage{eucal}
\usepackage{mathrsfs}
\usepackage[all]{xy}
\usepackage{epstopdf}
\usepackage{color}
\usepackage{float}
\usepackage{amsmath}
\usepackage{amsfonts}
\usepackage{bm}
\usepackage{natbib}
\usepackage{url}
\usepackage[colorlinks,urlcolor=blue]{hyperref}
\usepackage{setspace}
\usepackage[normalem]{ulem}
\usepackage{csvsimple}
\begin{document}

\title{Proof-of-concept for a nonadditive stochastic model of supercooled liquids}

\author{A. C. P. Rosa Jr.}\affiliation{Centro de Ci\^{e}ncias Exatas e das Tecnologias, Universidade Federal do Oeste da Bahia. Rua Bertioga, 892, Morada Nobre I, 47810-059 Barreiras, Bahia, Brazil.}
\author{E. Brito}\affiliation{Centro de Ci\^{e}ncias Exatas e das Tecnologias, Universidade Federal do Oeste da Bahia. Rua Bertioga, 892, Morada Nobre I, 47810-059 Barreiras, Bahia, Brazil.}
\author{W. S. Santana}\affiliation{Centro de Ci\^{e}ncias Exatas e das Tecnologias, Universidade Federal do Oeste da Bahia. Rua Bertioga, 892, Morada Nobre I, 47810-059 Barreiras, Bahia, Brazil.}
\author{C. Cruz\email{clebson.cruz@ufob.edu.br}}\affiliation{Centro de Ci\^{e}ncias Exatas e das Tecnologias, Universidade Federal do Oeste da Bahia. Rua Bertioga, 892, Morada Nobre I, 47810-059 Barreiras, Bahia, Brazil.}
\date{\today}

\begin{abstract}
The recently proposed non-additive stochastic model (NSM) offers a coherent physical interpretation for diffusive phenomena in glass-forming systems. This model presents non-exponential relationships between viscosity, activation energy, and temperature, characterizing the non-Arrhenius behavior observed in supercooled liquids. In this work, we fit the NSM viscosity equation to experimental temperature-dependent viscosity data corresponding to twenty-five glass-forming liquids and compare the fit parameters with those obtained using the Vogel-Fulcher-Tammann (VFT),  Avramov–Milchev (AM), and Mauro–Yue–Ellison–Gupta–Allan (MYEGA) models. The results demonstrate that the NSM provides an effective fitting equation for modeling viscosity experimental data in comparison with other established models (VFT, AM and MYEGA), characterizing the activation energy in fragile liquids, presenting a reliable indicator of the degree of fragility of the glass-forming liquids. 
\end{abstract}
\maketitle

\section{Introduction}
Glass science is a developing field of research due to the numerous technological applications of glass-forming liquids, and the formation mechanisms of amorphous solids remain an open question in the literature \cite{Berthier2011,Mauro2014.1,Mauro2014.2,PhysRevE.100.022139,Rosa2020}. The glass-forming process involves a liquid cooling to below its melting temperature without solidifying, resulting in a metastable state called supercooling. This process makes the fluid highly viscous, with molecular arrangement similar to the supercooled liquid near the glass transition \cite{Rosa2020,Berthier2011,Zheng2016}. In this scenario, Glass-forming liquids can be classified into two categories \cite{Zheng2016,PhysRevE.100.022139}
: strong liquids, which increase viscosity exponentially with temperature \cite{Truhlar2001}, and fragile liquids, which have non-exponential viscosity curves \cite{Truhlar2001,Rosa2016, Zheng2016,PhysRevE.100.022139}.

To get a satisfactory interpretation of the non-Arrhenius behavior in supercooled liquids, the authors of this paper developed the so-called nonadditive stochastic model (NSM) for the study of the reaction-diffusion processes in supercooled liquids \cite{PhysRevE.100.022139}. This proposed approach provides non-exponential functions for the thermal behavior of viscosity and the activation energy in fragile liquids.  
In other recent work, the group analyzed the viscosity equation in fragile liquids from the NSM, focusing on the relationship between activation energy and temperature \cite{Rosa2020}. The findings show that the fragility index is directly proportional to the activation energy for the glass transition temperature. In this regard, the developed approach appears in the literature as a robust method to characterize the degree of fragility of glass-forming systems and establishes the fragility index as a function of a $\gamma$ exponent \cite{Rosa2020,CarvalhoSilva2019,CarvalhoSilva2020,Kohout2021,Emran2022,TorregrosaCabanilles2022,Roy2022,Bondarchuk2023}.

In this context, this paper expands the previous research and serves as a proof-of-concept for the NSM in experimental settings. We analyze temperature-dependent viscosity data for twenty-five different glass-forming materials and use the NSM viscosity equation to determine the behavior of the activation energy with temperature. For each substance, we determine the glass transition temperature and the fragility index values. Additionally, we compare the NSM's fit parameters with those of other viscosity models such as the Vogel-Fulcher-Tammann (VFT), Avramov-Milchev (AM), and Mauro-Yue-Ellison-Gupta-Allan (MYEGA) models \cite{Mauro2009, Zhu2018}. The results demonstrate that the NSM is more accurate than these models for studying temperature-dependent viscosity in glass-forming liquids that exhibit super-Arrhenius behavior. Thus, the NSM provides a solid way to physically interpret the formation mechanisms of amorphous solids.

\section{Viscosity of glass-forming systems}

\begin{figure}[!htb]
	\centering
	\includegraphics[scale=0.5]{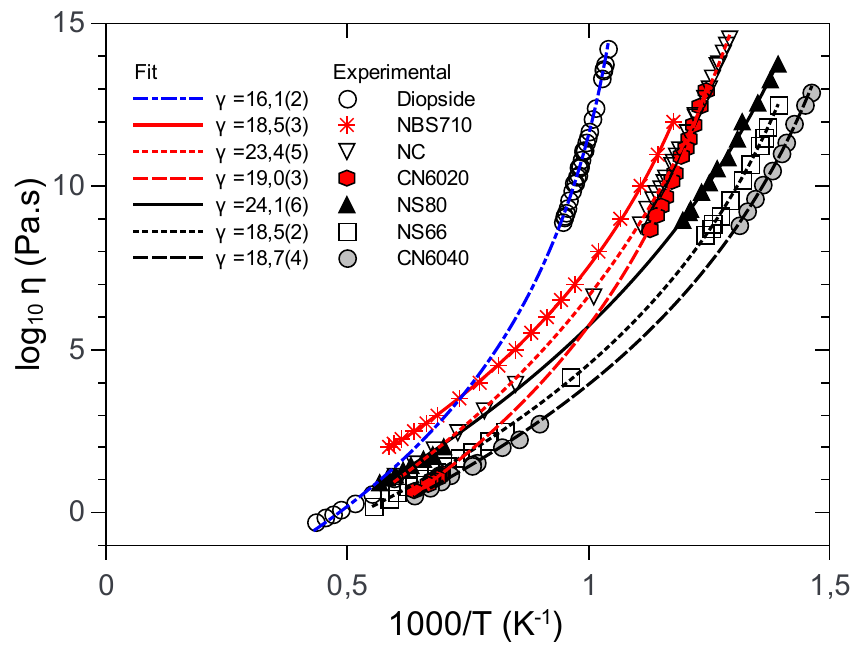}
	\caption{(Color online) Variation of the logarithm of viscosity as a function of the reciprocal temperature for silicate glasses \cite{Urbain1982,Sipp2001,Neuville2006,Jaccani2017}. The miscellaneous symbols corresponds to the experimental data, and the curve lines (continuous and dashed) to the fit of Eq.\ref{eq1}.}
	\label{fig:fig1}
\end{figure}

Viscosity is the reciprocal of fluidity \cite{Zheng2016,Rosa2020,PhysRevE.100.022139}, and the latter emerges naturally from the generalized drag coefficient that compounds the continuity equation in our formalism \cite{Rosa2020,PhysRevE.100.022139}. Thus, for the stationary regime of the reaction-diffusion process, the dependence of viscosity on temperature is characterized by a three-parameter model, written as,
\begin{equation}
\label{eq1}
\log_{10}\eta (T)=\log_{10}(\eta_{\infty})-\gamma \log_{10} \left[1-\frac{T_t}{T}\right],
\end{equation}
where $\eta_\infty$ is the high-temperature viscosity limit, $T_t$ is the viscosity divergence threshold temperature, and $\gamma$ is a characteristic exponent directly related to the fragility degree of the glass-forming liquid \cite{Rosa2020}. We validated the NSM by analyzing experimental temperature-dependent viscosity data from silicate glasses \cite{Urbain1982,Sipp2001,Neuville2006,Jaccani2017}, borosilicates \cite{Sipp1997}, aluminosilicates \cite{Urbain1982,Sipp2001,Gruener2001}, titania silicates \cite{Lika1996}, and chalcogenide glasses \cite{Kotl2010,Kotl2015,Zhu2018,Bartk2019}, totaling twenty-five glass-forming materials. Figure \ref{fig:fig1} shows the curves (continuous and dashed lines) corresponding to nonlinear regression fitting of the Eq.\ref{eq1} for viscosity experimental data (miscellaneous symbols) of the silicate glasses -- for other substances, see Figure \ref{fig:fig5} in Appendix \ref{appendix:A}.
Table \ref{tab1} (see Appendix \ref{appendix:A}) contains the fit parameters obtained for all glass-forming substances. The Pearson coefficient $R^2$ values are on the order of 0,999, and the $\chi^2$ test provides values below 0,1, demonstrating that Eq.\ref{eq1} provides an excellent fit to the viscosity-temperature data. From the $\gamma$ and $T_t$ values in Table \ref{tab1}, we calculated for all compositions the temperature-dependent activation energy, expressed by, 
\begin{equation}
\label{eq2}
E(T) =  \frac{\gamma R T_t}{1-\frac{T_t}{T}}
\end{equation}
where $R$ is the universal gas constant. Figure \ref{fig:fig2} shows the super-Arrhenius behavior for silicate glasses -- see Figure \ref{fig:fig6} in Appendix \ref{appendix:A} for the other substances. The activation energy is an increasing function of the reciprocal temperature, and $E(T)$ grows as it approaches the glass transition for all analyzed glass-forming materials. Considering the reference value $10^{12}$ Pa.s for viscosity \cite{Zheng2016} in Eq.\ref{eq1}, we can estimate the glass transition temperature $T_g$, given by,
\begin{equation}
\label{eq3}
	T_g = \frac{T_t}{1-10^{-B/\gamma}}
\end{equation}
where $B = 12 - log \eta_\infty$. Table \ref{tab1} (see Appendix \ref{appendix:A}) contains the $T_g$ values determined from Eq.\ref{eq3} for all glass-forming substances.
\begin{figure}[!htb]
	\centering
	\includegraphics[scale=0.5]{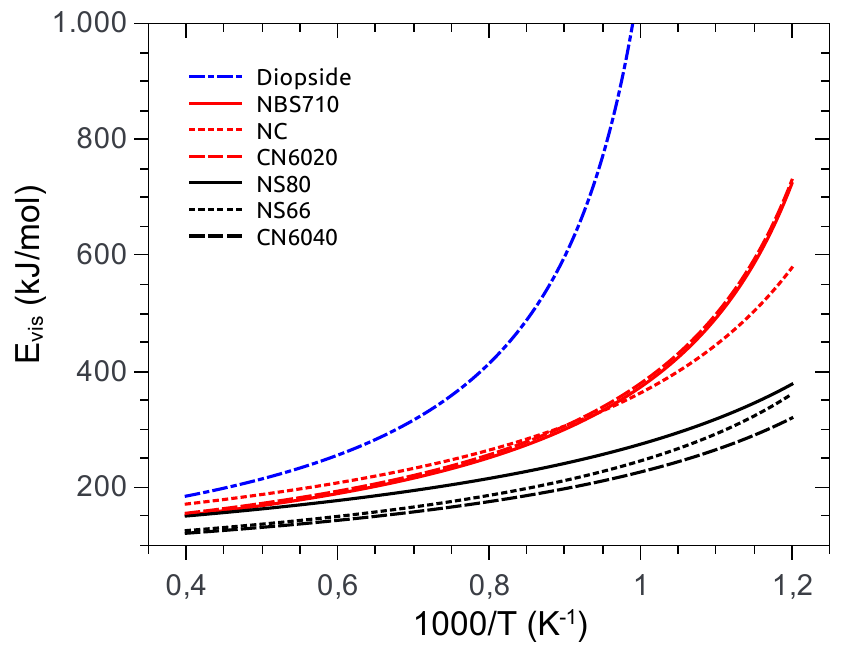}
	\caption{(Color online) Activation energy as a function of the reciprocal temperature using the fit parameters from Table \ref{tab1} in Eq.\ref{eq2}. Curve lines (continuous and dashed) for silicate glasses.}
	\label{fig:fig2}
\end{figure}
The NSM provides a reliable fragility index ($M_\eta$) directly proportional to $E_{vis}(T_g)$ \cite{Rosa2020} and, from Eq.\ref{eq1} and Eq.\ref{eq2}, we established one unique relation between the fragility index and the $\gamma$ exponent\cite{Rosa2020}, expressed as:
\begin{equation}
\label{eq4}
M_\eta = \gamma \left( 10^{B/\gamma} -1\right)
\end{equation}
  The concept of fragility in glass science is related to the degree of short-range order of the atomic arrangement of the supercooled liquid upon cooling through the glass transition so that, unlike strong liquids, a well-defined short-range order is lacking in fragile liquids \cite{Zhu2018}. According to Eq.\ref{eq4}, the greater the slope of the activation energy curve as the substance approaches the glass transition, the greater the fragility index value, which implies a lower value for the respective $\gamma$ exponent. Table \ref{tab1} (see Appendix \ref{appendix:A}) contains the $M_\eta$ values determined from Eq.\ref{eq4} for all glass-forming substances. Figure \ref{fig:fig3} shows the fragility index as a function of the $\gamma$ exponent, where the miscellaneous symbols correspond to the data in Table \ref{tab1} and the red line corresponds to Eq.\ref{eq4} for $B = 15$  ($log \eta_{\infty}  \approx -3$)\cite{Rosa2020}. The dispersion of the data relative to the theoretical curve is due to the variability of the experimental values of $log \eta_{\infty}$ obtained for each glass-forming substance. The horizontal line (dashed line) corresponds to the asymptotic limit between fragile and strong liquids, valid from the $\gamma \to \infty$ condition, implying $M_\eta \propto B$ (in \cite{Rosa2020} it was considered $M_\eta = B$ but, strictly speaking, we have that $M_\eta = B ln10$). 
\begin{figure}[!htb]
	\centering
	\includegraphics[scale=0.4]{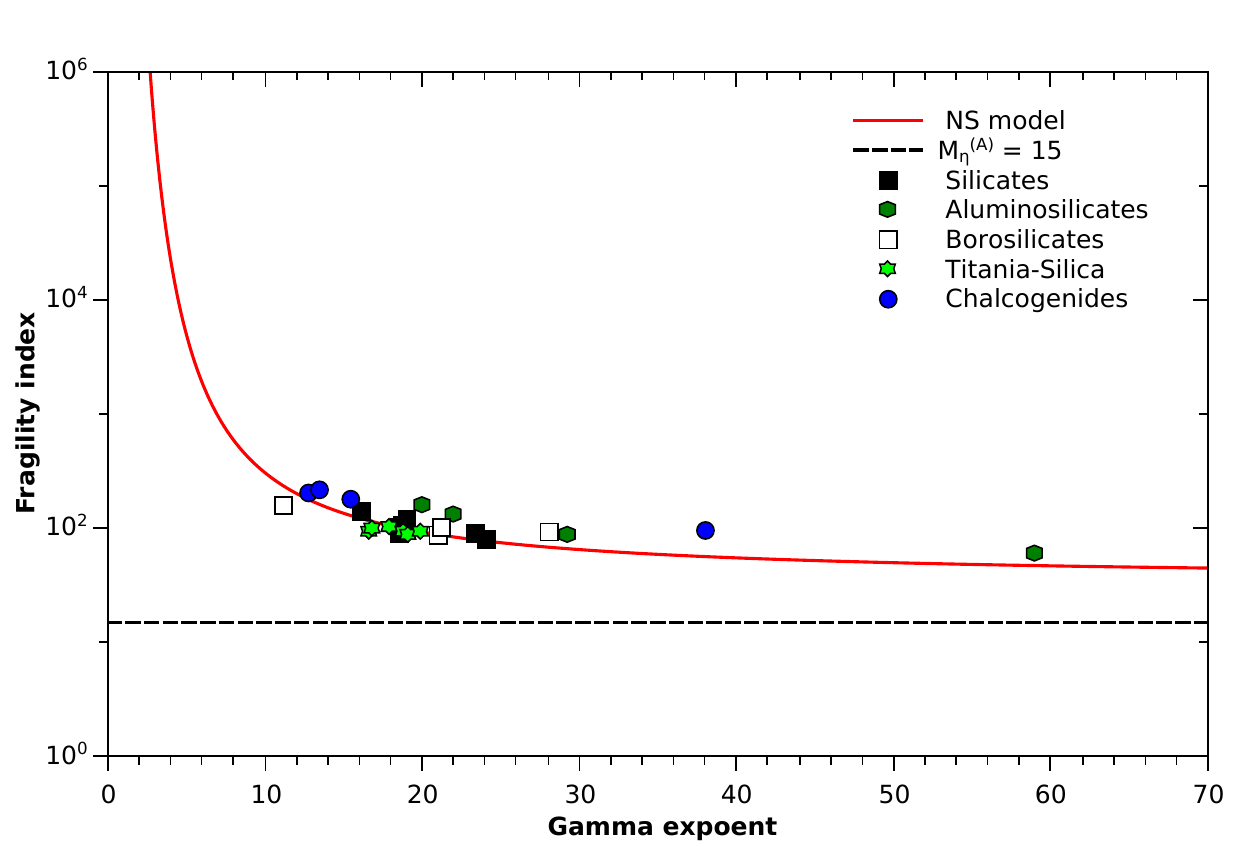}
	\caption{(Color online) The fragility index as a function of the exponent $\gamma$. The curve (red line) corresponds to the Eq.\ref{eq3} for $B=15$. The horizontal line (dashed line) corresponds to the value $M_\eta^{(A)} \approx 34,5$ and is an asymptotic limit between fragile and strong behaviors.}
	\label{fig:fig3}
\end{figure}

\section{viscosity models}
We compare the NSM with the Vogel-Fulcher-Tammann (VFT),  Avramov–Milchev (AM), and Mauro–Yue–Ellison–Gupta–Allan (MYEGA) models \cite{Mauro2009,Zhu2018}, which provide three efficient temperature-dependent viscosity equations to model non-Arrhenius behavior in supercooled liquids. We aim to test the accuracy of the NSM for fitting experimental data compared to other equations with three adjustment parameters but is not part of our study the analysis of the different theoretical interpretations that give rise to the four models. The VFT equation \cite{Zhu2018,Scherer1992} can be described as,
\begin{equation}
\label{eq5}
log_{10} \eta (T) = log_{10}(\eta_{\infty})+\frac{A}{T-T_0}
\end{equation}
where $\eta_\infty$, $B$, and $T_0$ are the fitting parameters. In this case, $A$ is an activation barrier, and $T_0$ is a viscosity divergence temperature. The AM model assumes that the structural disorder of the glass-forming liquid produces a random distribution of activation barriers whose dispersion around the mean activation energy depends on the system entropy \cite{Avramov1988}. The AM equation can be expressed as,
\begin{equation}
\label{eq6}
log_{10} \eta (T) = log_{10}(\eta_{\infty})+\left(\frac{\tau}{T}\right)^\alpha
\end{equation}
where $\eta_\infty$, $\tau$, and $\alpha$ are the fitting parameters.In this case, $\tau$ is a temperature associated with a reference entropy, and the exponent $\alpha$ is related to the fragility index that emerges from the AM model \cite{Zhu2018,Avramov1988}. The MYEGA equation, which derives from the Adam-Gibbs model that relates viscosity to the configurational entropy of the supercooled liquid \cite{Mauro2009}, is given by,
\begin{equation}
\label{eq7}
log_{10} \eta (T) = log_{10}(\eta_{\infty})+\frac{K}{T}exp\left(\frac{C}{T}\right)
\end{equation}
where $\eta_\infty$, $K$, and $C$ are the fitting parameters. In Eq.\ref{eq7}, the exponential function defines the configurational entropy of the supercooled liquid, and $K$ is an effective activation barrier. According to Table \ref{tab2} (see Appendix \ref{appendix:B}), the values of $\chi^2$ and $R^2$ obtained from the non-linear fit of the equations \ref{eq5}, \ref{eq6}, and \ref{eq7}, about the experimental data of all analyzed glass-forming substances, are in the same order of magnitude as the values obtained from Eq.\ref{eq1} for the same parameters. Therefore, the results demonstrate that the NSM provides as effective a fitting equation for modeling temperature-dependent viscosity experimental data as the VFT, AM, and MYEGA equations.  

\begin{figure}[!htb]
	\centering
	\includegraphics[scale=0.4]{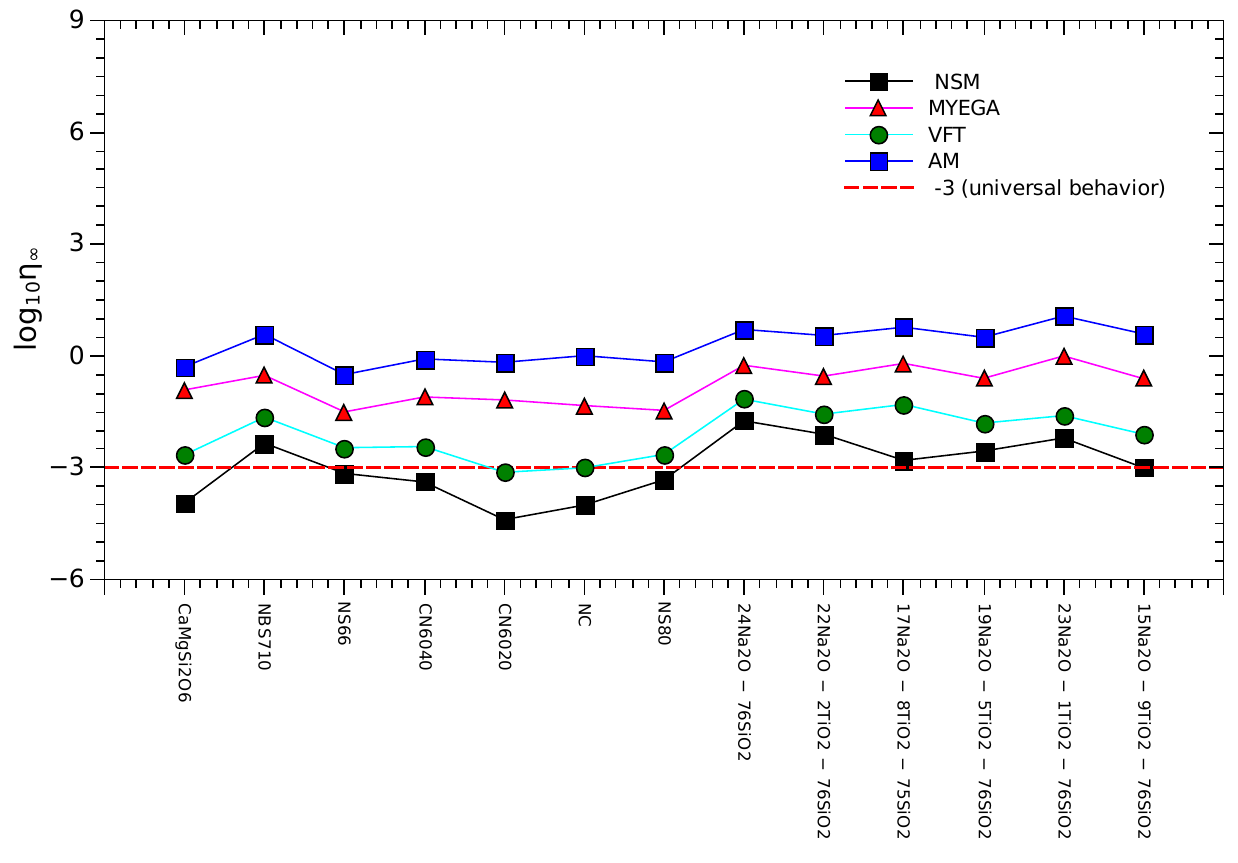}
	\caption{(Color online) The logarithm of the high-temperature viscosity limit for silicates \cite{Urbain1982,Sipp2001,Neuville2006,Jaccani2017} and titania silicates \cite{Lika1996} glasses. The miscellaneous symbols correspond to the values of the $log_{10} \eta_\infty$ obtained by the NSM (see Table \ref{tab1}), AM, VFT, and MYEGA (see Table \ref{tab2}) equations. The dashed red line corresponds to the universal behavior for $\eta_\infty$.}
	\label{fig:fig4}
\end{figure}

Table \ref{tab2} contains the $log_{10} \eta_\infty$ values obtained from the AM, VFT, and MYEGA models, given that the fit parameter is common to the four viscosity equations (see Table \ref{tab1} for the corresponding values of the NSM). Experimental evidence indicates that the condition $log_{10} \eta_\infty \approx -3$ is associated with universal behavior for the high-temperature viscosity limit \cite{Rosa2020,Zhu2018}, a result whose verification is beyond the scope of this work. Despite this, we can qualitatively discuss the results obtained for the high-temperature viscosity limit relative to the universal behavior mentioned above. Thus, Figure \ref{fig:fig4} illustrates the dispersion of the $log_{10} \eta_\infty$  fit values obtained by each of the four viscosity models (miscellaneous symbols) about the $-3$ reference value (red dashed line).

We considered only the silicate and titania-silicate glasses because the four viscosity equations provided less dispersed values for $log_{10}\eta_\infty$ for these materials if compared to other glass-forming liquids. The results demonstrated that the NSM provides $\eta_\infty$ values close to the limit of $10^{-3}$ Pa.s for silicate and titania-silica glasses, and other viscosity models tend to overestimate the high-temperature viscosity limit for the same substances, this behavior being more critical in the AM model.
\section{Conclusion}
 In summary, this work demonstrates a proof-of-concept for the NSM model in experimental settings.  The results demonstrated that the NSM is efficient for characterizing the temperature-dependent viscosity of glass-forming liquids in which the activation energy varies with the temperature. The values of $\chi^2$ and $R^2$ obtained from the non-linear regression of Eq.\ref{eq1} imply that the NSM accurately adjusted the experimental temperature-dependent viscosity data of the twenty-five glass-forming substances analyzed. These parameters are in the same order of magnitude if compared to values of the $\chi^2$ and $R^2$ obtained by the VFT, AM, and MYEGA equations, widely used to model temperature-dependent viscosity in supercooled liquids. From the fit parameters of Eq\ref{eq1}, we calculated the experimental values from the activation energy (see Eq.\ref{eq2}) and the glass transition temperature (see Eq.\ref{eq3}). Also, we demonstrated the robustness of Eq.\ref{eq4}, what makes the exponent $\gamma$ a reliable indicator of the degree of fragility of the glass-forming substance. Finally, we verified that, while the NSM provided values for the high-temperature viscosity limit close to the universal behavior $10^{-3}$ Pa.s from the silicates and titania-silica glasses, the other viscosity models overestimate $\eta_\infty$ for the same substances. Thus, the results demonstrate that the NSM guarantees a solid interpretation for diffusive processes in supercooled liquids that exhibit non-Arrhenius behavior and consolidates the NSM as a robust theoretical basis for the physical interpretation of the dynamic properties of glass-forming systems.

\begin{acknowledgments}
The authors would like to thank the Bahia State Research Support Foundation (FAPESB) for its financial support.
\end{acknowledgments}

\appendix
\section{Fit parameters of the NSM}
\label{appendix:A}
\begin{figure*}[ht!]
	\centering
	\subfigure[]{\includegraphics[scale=0.5]{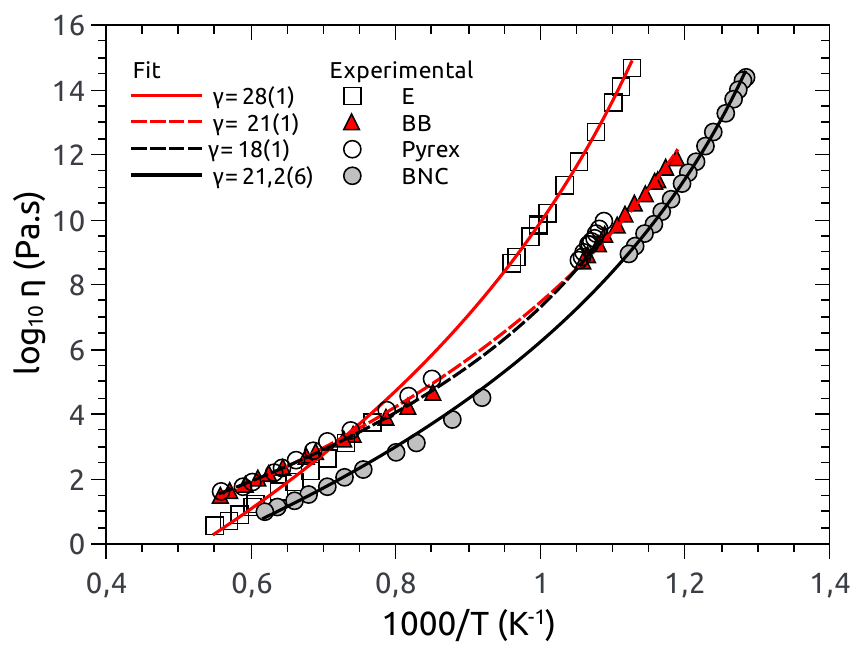}}
	\subfigure[]{\includegraphics[scale=0.5]{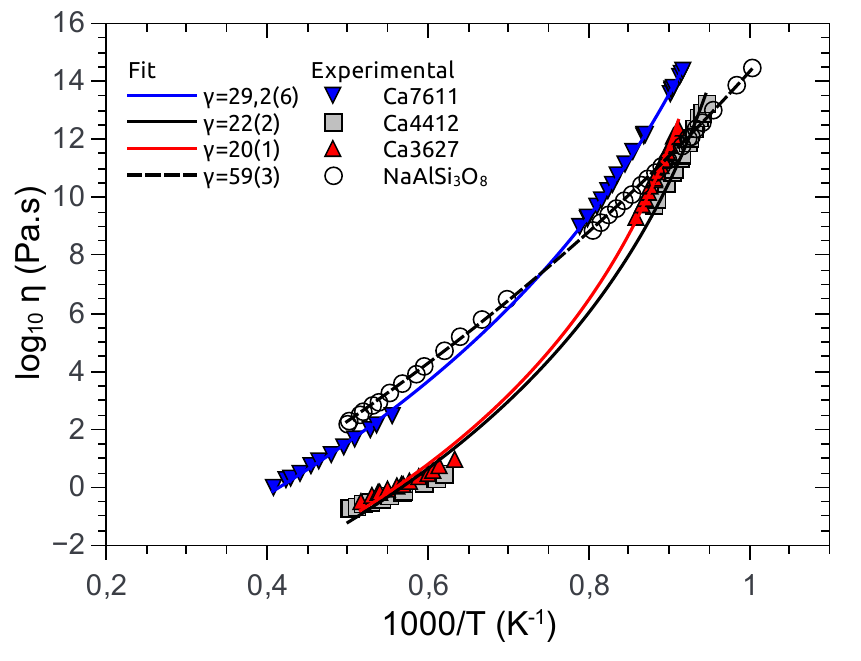}}\\
	\subfigure[]{\includegraphics[scale=0.5]{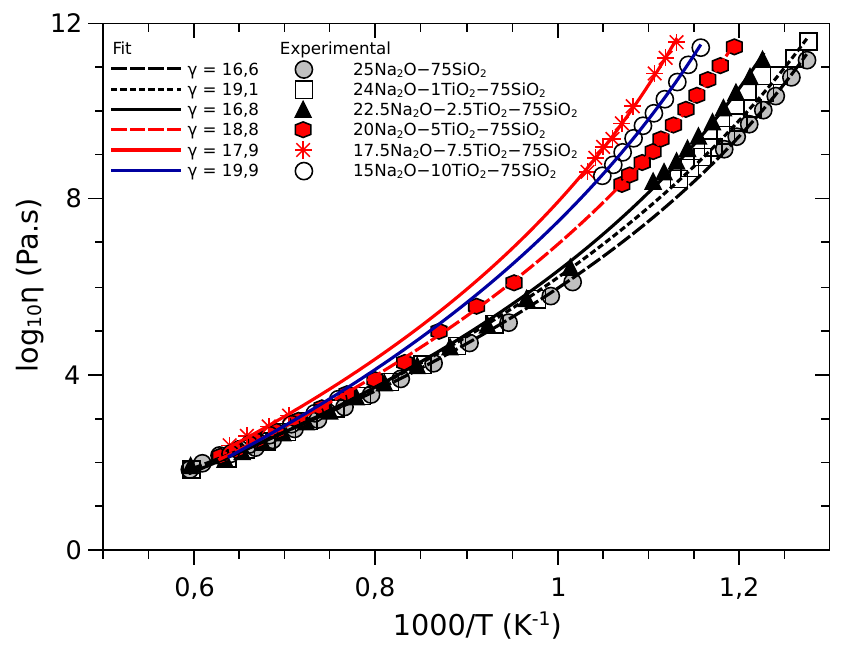}}
	\subfigure[]{\includegraphics[scale=0.5]{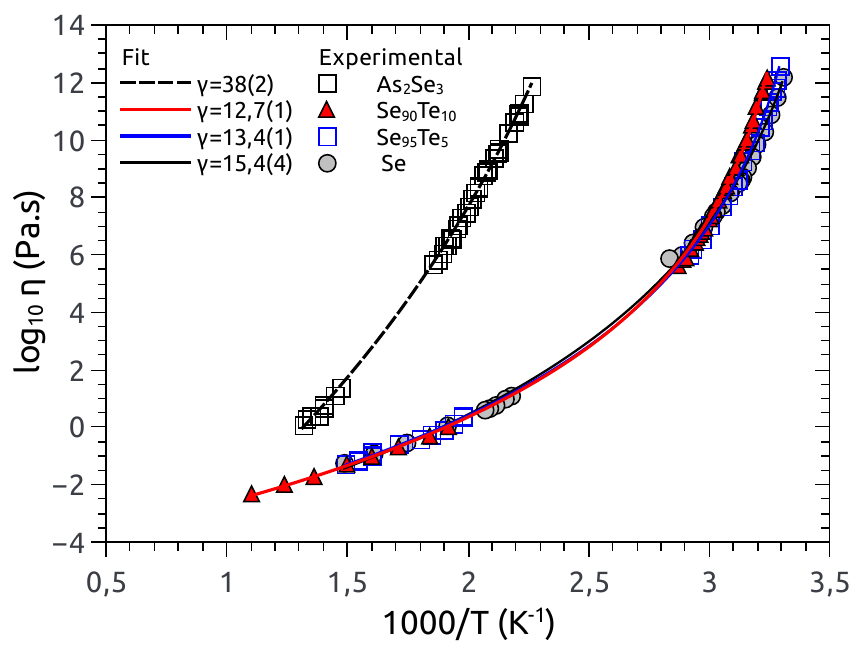}}
	\caption{(Color online) Variation of the logarithm of viscosity as a function of the reciprocal temperature for glass-forming liquids. Experimental data (miscellaneous symbols):  (a) borosilicates\cite{Sipp1997}, (b) aluminosilicates\cite{Urbain1982,Sipp2001,Gruener2001}, (c)  titania silicates\cite{Lika1996}, and (d) chalcogenides \cite{Kotl2010,Kotl2015,Zhu2018,Bartk2019} glasses. Curve lines (continuous and dashed): fit the Eq.\ref{eq1} for (a)  borosilicates, (b) aluminosilicates, (c) titania silicates, and (d) chalcogenides glasses.} 
	\label{fig:fig5}
\end{figure*}

\begin{figure*}[ht!]
	\centering
	\subfigure[]{\includegraphics[scale=0.5]{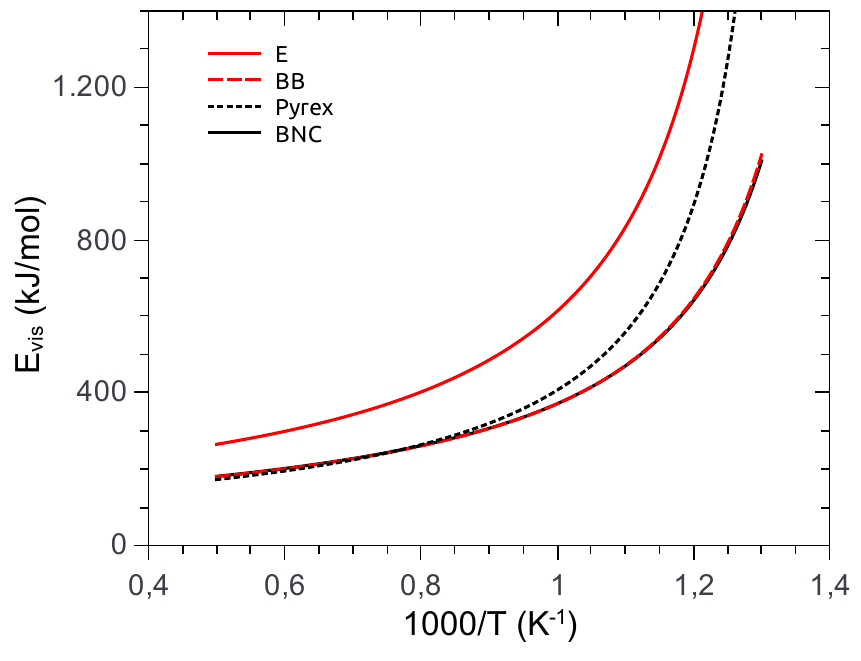}}
	\subfigure[]{\includegraphics[scale=0.5]{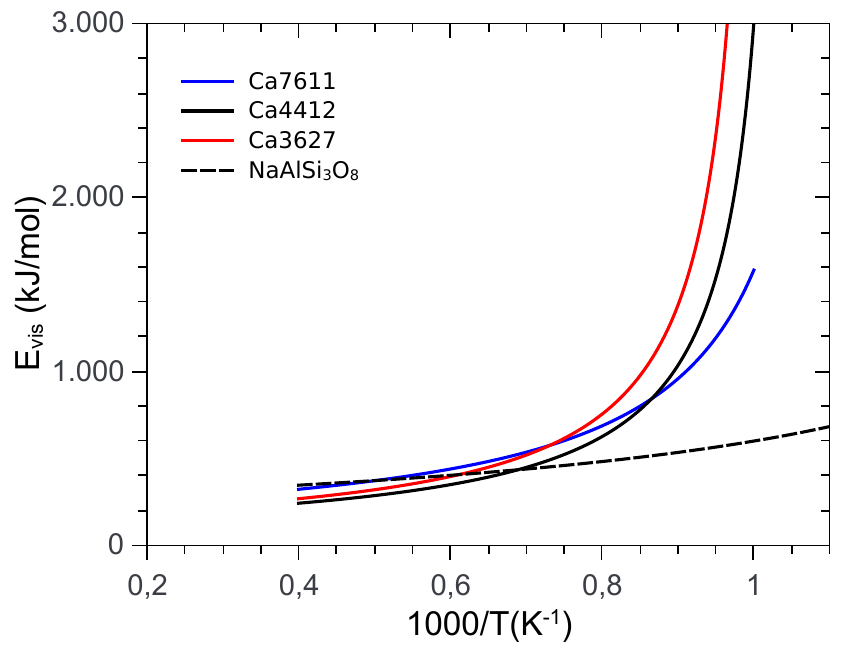}}\\
	\subfigure[]{\includegraphics[scale=0.5]{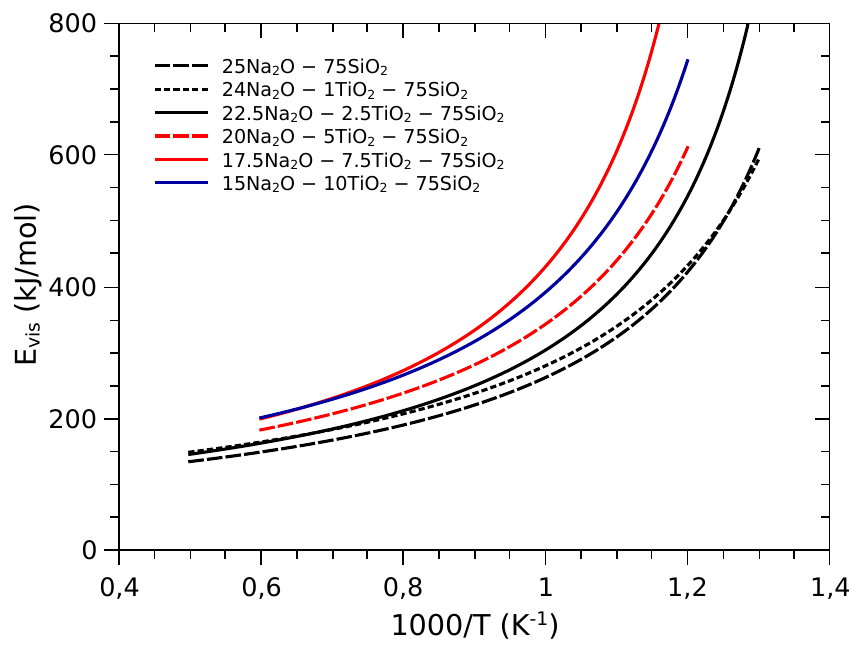}}
	\subfigure[]{\includegraphics[scale=0.5]{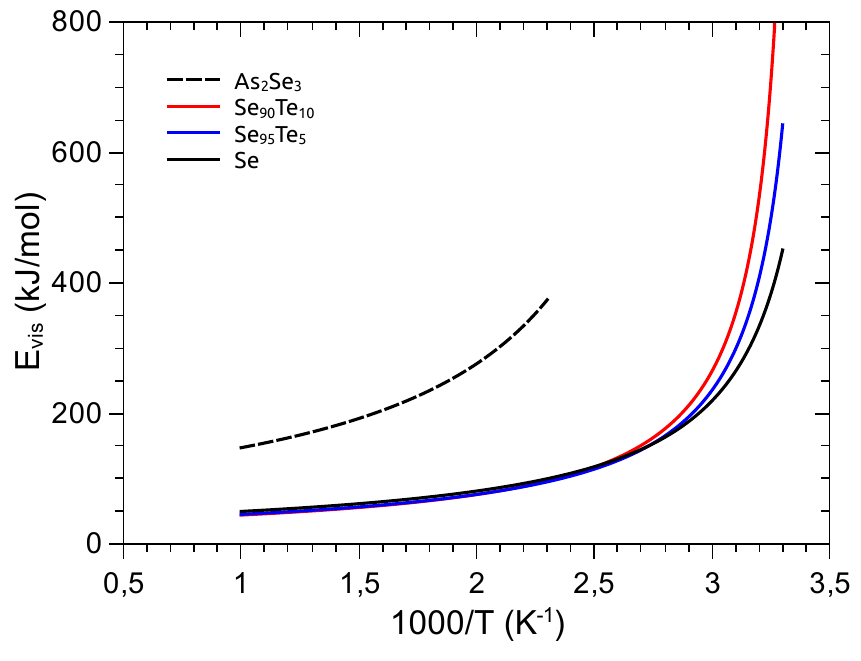}}
	\caption{(Color online) Activation energy as a function of the reciprocal temperature using the fit parameters from Table \ref{tab1} in Eq.\ref{eq2}. Curve lines (continuous and dashed) for (a) silicates, (b) borosilicates, (c) aluminosilicates, (d) titania silicates, and (e) chalcogenides glasses.}
	\label{fig:fig6}
\end{figure*}

Similar to Figure \ref{fig:fig1}, Figure \ref{fig:fig5} shows the logarithm of viscosity as a function of the reciprocal temperature for other glass-forming substances, where the miscellaneous symbols are experimental viscosity data refer to (5a) borosilicates \cite{Sipp1997}, (5b) aluminosilicates \cite{Urbain1982,Sipp2001,Gruener2001}, (5c) titania silicates\cite{Lika1996}, and (5d) chalcogenides glasses. The continuous and dashed lines correspond to the nonlinear regression fitting of Eq.\ref{eq1} using the Levenberg-Marquadt method \cite{Bellavia2018} for all cases.  Table \ref{tab1} contains the fit parameters $\gamma$, $log_{10}\eta_\infty$ and $T_t$ from the Eq.\ref{eq1}, the Pearson coefficient $R^2$ and the $\chi^2$ test obtained for all glass-forming substances analyzed.  The fragility index values calculated by Eq.\ref{eq3} and the glass transition temperature values by Eq.\ref{eq4} complete Table \ref{tab1}. We apply the error propagation method to determine the uncertainties about the $M_\eta$ and $T_g$ values. Similar to Figure \ref{fig:fig2}, Figure \ref{fig:fig6} shows the activation energy as a function of the reciprocal temperature for (6a) borosilicates, (6b) aluminosilicates, (6c) titania silicates, and (6d) chalcogenides glasses. Curve lines (continuous and dashed) correspond to Eq.\ref{eq2} using the fit parameters from Table \ref{tab1}. 

\begin{table*}[!]
	\caption{Fit parameters of the Eq.\ref{eq1} for temperature-dependent viscosity data of twenty-five glass-forming liquids, including silicates\cite{Urbain1982,Sipp2001,Neuville2006,Jaccani2017}, borosilicates\cite{Sipp1997}, aluminosilicates\cite{Urbain1982,Sipp2001,Gruener2001}, titania silicates\cite{Lika1996}, and chalcogenides\cite{Kotl2010,Kotl2015,Zhu2018,Bartk2019} glasses. We calculated the $T_g$ values by the Eq.\ref{eq3} and  the $M_\eta$ values by the Eq.\ref{eq4}.}
	\centering
	\resizebox{\textwidth}{!}{
		\begin{tabular}{ccccccccc}
			\noalign{\smallskip} \hline \hline \noalign{\smallskip}
			Glass&$\log_{10}\eta_\infty$ & $R^2$ & $\chi^2$ &$\gamma$& $T_t$ (K) & $T_g$ (K) & $M_\eta$\\
			&&&($\times 10^{-2}$)&&&&\\
			\hline 
			Silicates&&&&&&&\\
			\hline
			$CaMgSi_2O_6$&-3,94(7)&0,9994&1,1&16,1(2)&890(3)&991(22)&141(4)\\
			NBS710&-2,34(5)&0,9999&0,1&18,5(3)&708(3)&851(23)&92(3)\\
			NS66&-3,15(4)&0,9999&0,2&18,5(2)&615(2)&725(12)&103(2)\\
			CN6040&-3,38(7)&0,9998&0,4&18,7(4)&593(4)&698(21)&106(4)\\
			CN6020&-4,39(6)&0,9999&0,3&19,0(3)&706(3)&818(17)&119(3)\\
			NC&-4,0(1)&0,9996&0,8&23,4(5)&651(4)&821(27)&90(4)\\
			NS80&-3,32(8)&0,9998&0,5&24,1(6)&578(5)&752(27)&80(4)\\
			\hline 
			Aluminosilicates&&&&&&&\\
			\hline
			Ca3627&-7,0(4)&0,9984&5,1&20(1)&980(20)&1104(87)&158(15)\\
			Ca4412&-6,5(3)&0,9976&9,2&22(2)&950(10)&1110(114)&131(18)\\
			Ca7611&-5,65(8)&0,9998&0,6&29,2(6)&867(6)&1154(30)&88(3)\\
			$NaAlSi_3O_8$&-5,9(1)&0,9998&0,3&59(3)&550(10)&1094(62)&60(5)\\
			\hline 
			Borosilicates&&&&&&&\\		
			\hline
			Pyrex&-1,2(2)&0,9977&3,1&11,1(6)&832(8)&890(156)&160(30)\\
			BB&-2,9(2)&0,9992&1,4&21(1)&680(12)&845(75)&87(9)\\
			BNC&-4,2(1)&0,9994&1,6&21,2(6)&678(5)&819(31)&102(5)\\
			E&-5,9(2)&0,9994&1,7&28(1)&725(9)&941(48)&94(6)\\
			\hline 
			Titania-Silica&&&&&&&\\
			\hline
			$24 Na_{2}O-76 SiO_{2}$&-1,74(7)&0,9998&0,3&16,6(4)&655(6)&769(37)&95(5)\\
			$22 Na_2O-2 TiO_2-76 SiO_2$&-2,1(1)&0,9994&0,7&16,8(6)&685(7)&801(48)&99(7)\\
			$17 Na_2O-8 TiO_2-75 SiO_2$&-2,8(1)&0,9998&0,2&17,9(5)&743(6)&873(40)&102(6)\\
			$19 Na_2O-5 TiO_2-76 SiO_2$&-2,55(9)&0,9998&0,2&18,8(4)&687(5)&826(35)&93(4)\\
			$23 Na_2O-1 TiO_2-76 SiO_2$&-2,20(8)&0,9998&0,3&19,1(5)&638(6)&779(36)&87(5)\\
			$15 Na_2O-9 TiO_2-76 SiO_2$&-3,0(1)&0,9998&0,3&19,9(7)&703(8)&853(42)&93(6)\\
			\hline 
			Chalcogenides&&&&&&&\\
			\hline
			$Se_{90}Te_{10}$&-3,7(1)&0,9996&0,7&12,7(1)&296,6(8)&315(9)&206(6)\\
			$Se_{95}Te_{5}$&-4,75(5)&0,9998&0,7&13,4(1)&287,9(5)&306(4)&218(3)\\
			Se&-5,0(1)&0,9988&2,4&15,4(4)&279(2)&303(10)&180(8)\\
			$As_2Se_3$&-8,9(2)&0,9993&1,0&38(2)&318(6)&443(27)&97(8)\\
			\hline 
		\end{tabular}
	}
	\label{tab1}
\end{table*}
\section{Fit paramaters of the AM, VFT, and MYEGA models}
\label{appendix:B}

Table \ref{tab2} list the values of $\log_{10}\eta_\infty$, $R^2$, and $\chi^2$ obtained from the non-linear regression fitting  of the  Eq.\ref{eq5} (VFT), Eq.\ref{eq6} (AM), and Eq.\ref{eq7} (MYEGA) using the Levenberg-Marquadt method \cite{Bellavia2018} for experimental temperature-dependent viscosity data of the twenty-five glass-forming liquids analyzed in this work.
\begin{table*}[ht!]
	\centering
	\caption{{$\log_{10}\eta_\infty$, $R^2$, and $\chi^2$ obtained from AM, VFT, and MYEGA equations for temperature-dependent viscosity data of twenty-five glass-forming liquids, including silicates\cite{Urbain1982,Sipp2001,Neuville2006,Jaccani2017}, borosilicates\cite{Sipp1997}, aluminosilicates\cite{Urbain1982,Sipp2001,Gruener2001}, titania silicates\cite{Lika1996}, and chalcogenides\cite{Kotl2010,Kotl2015,Zhu2018,Bartk2019} glasses.}}
	\resizebox{\textwidth}{!}{
		\begin{tabular}{cccccccccc}
			\noalign{\smallskip} \hline \hline \noalign{\smallskip}
			Glass&&AM&&&VFT&&&MYEGA&\\ 
			&$\log_{10}\eta_\infty$ & $R^2$ & $\chi^2$ & $\log_{10}\eta_\infty$ & $R^2$ & $\chi^2$ & $\log_{10}\eta_\infty$ & $R^2$ & $\chi^2$\\
			&&&($\times 10^{-2}$)&&&($\times 10^{-2}$)&&&($\times 10^{-2}$)\\
			\hline 
			Silicates&&&&&&&&&\\
			\hline
			$CaMgSi_2O_6$&-0,29(4)&0,9996&0,8&-2,65(4)&0,9999&0,3&-0,91(3)&0,9999&0,3\\
			NBS710&0,58(8)&0,9997&0,3&-1,642(7)&0,99999&0,002&-0,52(6)&0,9999&0,1\\
			NS66&-0,5(2)&0,9988&2,7&-2,46(8)&0,9998&0,5&-1,5(2)&0,9993&1,5\\
			CN6040&-0,08(14)&0,9994&1,6&-2,43(6)&0,9999&0,2&-1,1(1)&0,9997&0,7\\
			CN6020&-0,17(4)&0,99993&0,2&-3,12(5)&0,9999&0,1&-1,18(5)&0,99996&0,1\\
			NC&0,007(72)&0,9998&0,4&-3,00(7)&0,9998&0,3&-1,33(6)&0,9999&0,2\\
			NS80&-0,16(8)&0,9998&0,6&-2,64(8)&0,9996&0,4&-1,46(9)&0,9999&0,4\\
			\hline 
			Aluminosilicates&&&&&&&&&\\
			\hline
			Ca3627&-1,6(1)&0,9998&0,7&-5,1(3)&0,9991&3&-2,6(2)&0,9996&1,0\\
			Ca4412&-2,0(1)&0,9995&1,8&-5,5(4)&0,9984&6&-3,0(3)&0,9992&3,0\\
			Ca7611&-1,59(6)&0,99991&0,3&-4,75(6)&0,9999&0,3&-3,23(5)&0,99997&0,09\\
			$NaAlSi_3O_8$&-3,1(3)&0,9997&0,5&-5,5(2)&0,9998&0,3&-5,1(2)&0,9998&0,4\\
			\hline 
			Borosilicates&&&&&&&&&\\		
			\hline
			Pyrex&1,0(4)&0,9931&9,6&-4,5(3)&0,9959&5,7&0,4(5)&0,9940&8,3\\
			BB&0,11(9)&0,9998&0,4&-2,2(2)&0,9995&0,9&-1,1(1)&0,9997&0,5\\
			BNC&0,005(51)&0,9999&0,3&-2,99(9)&0,9998&0,5&-1,16(3)&0,99998&0,06\\
			E&-1,4(3)&0,99997&0,06&-4,8(1)&0,9997&0,8&-3,04(8)&0,99992&0,2\\
			\hline 
			Titania-Silica&&&&&&&&&\\
			\hline
			$24 Na_{2}O-76 SiO_{2}$&0,71(8)&0,9997&0,3&-1,16(6)&0,9999&0,1&-0,25(7)&0,9999&0,2\\
			$23 Na_2O-1 TiO_2-76 SiO_2$&0,55(8)&0,9998&0,3&-1,56(6)&0,9999&0,1&-0,54(7)&0,9999&0,1\\
			$22 Na_2O-2 TiO_2-76 SiO_2$&0,77(9)&0,9996&0,4&-1,3(1)&0,9997&0,4&-0,2(1)&0,9997&0,3\\
			$19 Na_2O-5 TiO_2-76 SiO_2$&0,5(1)&0,9997&0,4&-1,80(8)&0,9999&0,2&-0,6(1)&0,9998&0,2\\
			$17 Na_2O-8 TiO_2-75 SiO_2$&1,07(8)&0,9999&0,2&-1,6(1)&0,9999&0,2&0,0002(103)&0,99992&0,1\\
			$15 Na_2O-9 TiO_2-76 SiO_2$&0,59(4)&0,99998&0,03&-2,10(9)&0,99992&0,1&-0,61(5)&0,99999&0,02\\
			\hline 
			Chalcogenides&&&&&&&&&\\
			\hline
			$Se_{90}Te_{10}$&-1,5(1)&0,9964&7,6&-3,35(5)&0,9997&0,7&-1,9(1)&0,9977&5,0\\			
			$Se_{95}Te_{5}$&-1,0(1)&0,9978&6,2&-3,20(6)&0,9996&1&-1,5(1)&0,9986&4,0\\
			$Se$&-1,3(2)&0,9968&6,3&-3,7(2)&0,9985&3&-2,1(2)&0,9976&5,0\\
			$As_2Se_3$&-4,1(2)&0,9995&0,6&-8,0(2)&0,9994&0,9&-6,4(3)&0,9994&0,7\\
			\hline 
		\end{tabular}
	}
	\label{tab2}
\end{table*}

\end{document}